\documentclass[aps,pre,superscriptaddress,twocolumn]{revtex4-1}
\usepackage{graphicx}                   
\usepackage{hyperref}                   
\usepackage{amsmath,amssymb}            
\usepackage{multirow}		
\usepackage{xcolor}	
\usepackage{amsmath,color}

\graphicspath{{figs/}}					

\begin{document}

\author{Mohadeseh Feshanjerdi}\email{m.feshanjerdi@alzahra.ac.ir}
\affiliation{Department of Physics, University of Tehran, P. O. Box 14395-547, Tehran, Iran}
\author{Abbas Ali Saberi}\email{(corresponding author) ab.saberi@ut.ac.ir}
\affiliation{Department of Physics, University of Tehran, P. O. Box 14395-547, Tehran, Iran}
\affiliation{Institut f\"ur Theoretische
  Physik, Universit\"at zu K\"oln, Z\"ulpicher Str. 77, 50937 K\"oln,
  Germany}

\title{Universality class of epidemic percolation transitions driven by random walks}


\begin{abstract}
Inspired by the recent viral epidemic outbreak and its consequent worldwide pandemic, we devise a model to capture the dynamics and the universality of the spread of such infectious diseases. The transition from a pre-critical to the post-critical phase is modeled by a percolation problem driven by random walks on a two-dimensional lattice with an extra average number $\rho$ of nonlocal links per site. Using the finite-size scaling analysis, we find that the effective exponents of the percolation transitions as well as the corresponding time thresholds, extrapolated to the infinite system size, are $\rho$-dependent. We argue that the $\rho$-dependence of our estimated exponents represents a crossover-type behavior caused by the finite-size effects between the two limiting regimes of the system. We also find that the universal scaling functions governing the critical behavior in every single realization of the model, can be well described by the theory of extreme values for the maximum jumps in the order parameter and by the central limit theorem for the transition threshold.
\end{abstract}

\maketitle


\section{Introduction}
Percolation is one of the simplest models in probability theory which provides a suitable platform to formulate and model various natural phenomena that exhibit geometric phase transitions with universal characteristics~\cite{r1,r2, r2-1,r2-2}. 
Percolation has been applied to describe a wide range of critical behavior such as, among many others, flow through porous media for connectivity \cite{r3,r4}, networks \cite{r5,r6,r7}, magnetic
models \cite{r8, r9, r10, r11, r11-1,r11-2, r11-3, r11-4}, colloids \cite{r12, r13}, growth models \cite{r14}, topography of planets \cite{r14-1, r14-2, r14-3} and epidemic models \cite{r14-4}. The concept of universality lets many microscopically different physical systems exhibit the same critical behavior with quantitatively identical features, assigned by a set of critical exponents. Thanks to the emergence of the powerful conformal field theories (CFTs) in two dimensions (2D), many exact results have been obtained for the percolation model in 2D \cite{r14-5}. 

On the other hand, the theory of random walks, also known as Brownian motion in the scaling limit, plays an essential role in many stochastic processes and challenging problems in probability and statistical physics \cite{r14-6, r14-7, r14-8}. In two dimensions, random walks and percolation at its critical point share various interesting properties other than their conformal invariant property. For example, the external perimeters of a critical percolation cluster and a planar random walk in the continuum limit are believed to be described by the scaling limit of planar self-avoiding random walks (SAWs) \cite{r14-9}. Moreover, the trace of a smart kinetic self-avoiding random walk \cite{r22} (i.e., the trace of a random walker that never crosses itself and never gets trapped) produces statistically the same fractal path in accord with the perimeter (or the hull) of a critical percolation cluster in 2D.

Here we combine these two fundamental theories with some modifications to develop a model to study the universal behavior and dynamics of the spread of infectious diseases over a 2D square lattice possessing a variable percentage $\rho$ of nonlocal extra links. In the absence of the nonlocal links i.e., the trivial case with $\rho=0$, there will be a random walker (infected person) which randomly visits the lattice sites (the people prone to disease) and the average linear size of the visited (infected) sites grows with the number $N$ of walks as $\sim \sqrt N$. For a square lattice of linear size $L$ and periodic boundary conditions, this means that the percolation of the infected sites is expected to happen after $t_c=N\sim L^2$ steps which diverges when the system size $L$ goes to infinity. In the following, we will indeed study the dynamics of the model on square lattices with nonzero $\rho$.\\
We use finite-size scaling (FSS) hypothesis \cite{r22-1} to precisely locate the critical thresholds  in the limit of the infinite system size $L\rightarrow\infty$ and extract various critical exponents of the model for $\rho>0$. Instead of performing a traditional FSS analysis at the critical
phase transition point, we rather follow the universal framework of a FSS analysis recently developed in \cite{r26} based on the statistics and critical scaling of the size of the largest gap in the order parameter. The key-point of \cite{r26} is the following: Since the size of the largest gap (within a time-series of the evolving largest cluster of the infected sites) can be regarded as an extreme value of the percolation process, the corresponding scaling function is believed to be governed by extreme-value statistics. Such intuition has led to unify continuous and discontinuous percolation transitions by identifying the universal critical scaling functions as the extreme-value Gumbel distribution \cite{r26}. This framework also provides an alternative approach to estimate various critical exponents and fractal dimensions which determine the universality class and the type (continuous or discontinuous) of the transition.

This paper is structured as follows: The following Section \ref{Model} gives the details of our devised model and the dynamics. Section \ref{observbls} goes over the theoretical backgrounds and definition of various percolation observables that we are going to measure. Section \ref{numerics} presents our numerical results and their comparison with our analytical arguments. A full list of the effective critical exponents estimated for finite-size systems for various choices of $\rho$ is also summarized. The final Section \ref{concl} concludes and provides suggestions for future research. 

\section{The Model}\label{Model}

Consider a square lattice of size $L\times L$ with periodic boundary conditions in both directions. Initially, each of the $N=L^2$ sites (or susceptible individuals) has four nearest neighbors which are connected by local links of a lattice constant $a=1$ length. For a given $\rho=$(number of extra nonlocal links)/$N$, we select a pair of randomly chosen sites and connect them together by an extra nonlocal link which effectively makes them nearest neighbors. This procedure continues until the desired value $\rho$ is reached. The lattice with the only parameter $\rho$ is now ready to study our stochastic epidemic modeling. 

\begin{figure}[t]
	\centering
	\includegraphics[width=3.3in]{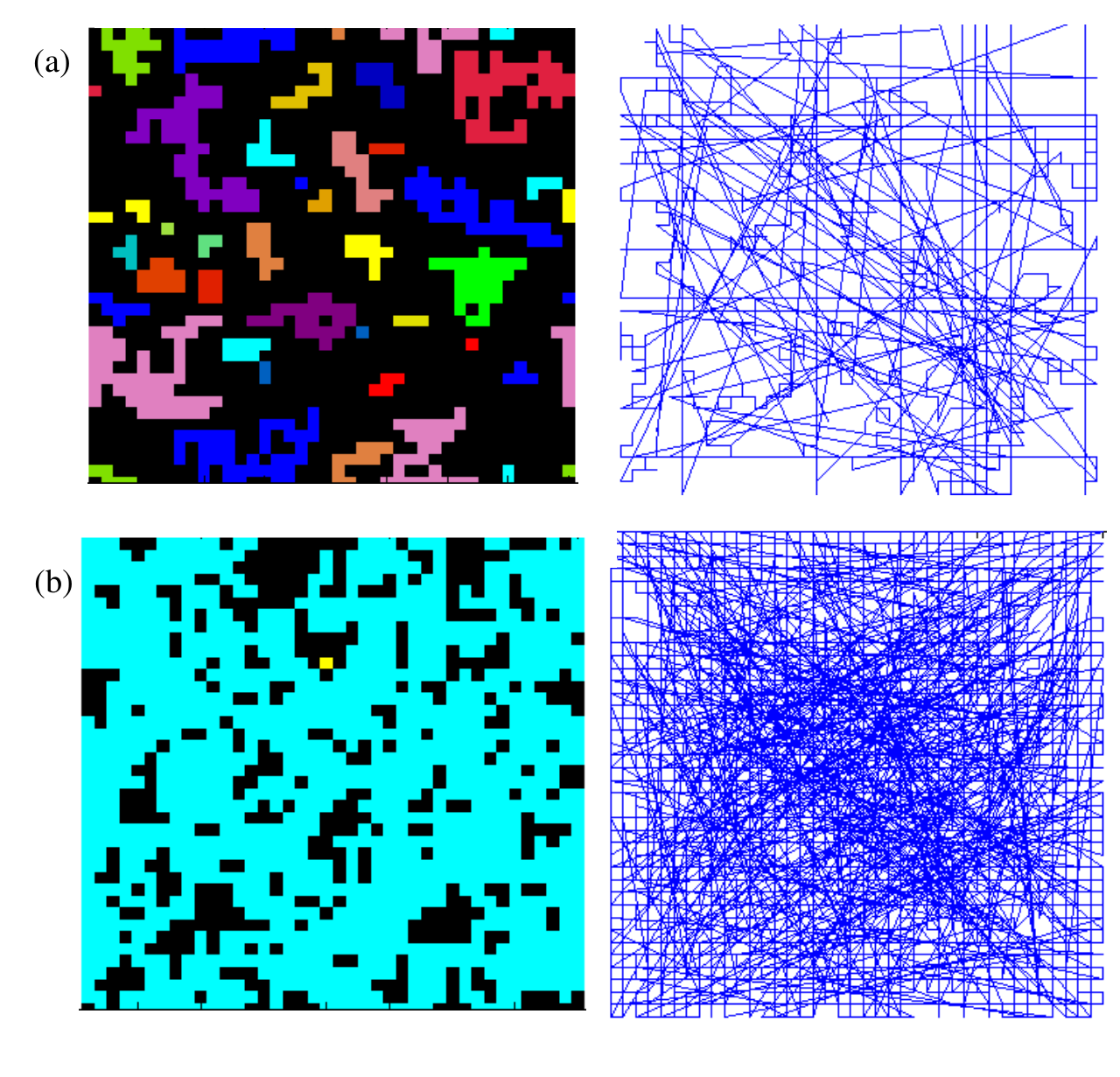}
	\caption{Illustration of the infected clusters in the epidemic outbreak (left-up panel) and pandemic (left-down panel) phases at time $t<t_c$ and $t>t_c$, respectively, on a square lattice of linear size $L=40$ and $\rho=0.8$ with periodic boundary conditions in both directions. Every disjoint infected cluster is shown by a different color. Uninfected sites are shown in black. The corresponding steps taken by the random walker in the time interval [0, $t$] for each phase are shown in the right panels. }
	\label{figure-1}
\end{figure}

To initiate the dynamics of our model, a random walker (or the infected individual) starts its journey from a randomly chosen site on the lattice at $t=0$. For the ease of notation in the remainder of this paper, we refer to time $t$ as the number of random walk steps over $N$, i.e., $t:=t/N$. At each time step, the random walker moves to one of its randomly chosen nearest neighboring sites (either the local or nonlocal ones). Every visited site by the random walker is marked as an infected site upon visiting (no matter how many times it is visited) and remains infected forever. A cluster of infected sites is identified as a set of nearest neighboring infected sites which are connected only by local links of length $a=1$. At early times $t<t_c$, there will only appear small-size disjoint clusters all over the lattice (epidemic outbreak phase) and they gradually merge into a single super cluster at $t>t_c$ which spans the entire lattice (pandemic phase). Figure \ref{figure-1} illustrates these two phases (left panels) and steps taken in each phase (right panels) on a lattice of size $L=40$ and $\rho=0.8$. Characterizing the nature of this phase transition at $t=t_c$ and its universality class by determining the corresponding critical exponents as well as the critical time $t_c$ where the epidemic prevalence peaks, are the main subjects of our present study. 

\section{theoretical background and Definition of the observables} \label{observbls}

The relative size $s(t)$ of the largest cluster at time $t$ is known as the standard order parameter in percolation transitions. Once the random walker starts moving, the infected clusters form and evolve in time. We monitor the evolution of $s(t)$ at every time step $\Delta t$, and record its maximum jump,
\begin{equation}\label{1} 
\Delta \equiv \max_t[s(t+\Delta t)-s(t)],
\end{equation}
and the time of incidence $t_c$, for every single realization. We also record the size of the largest cluster just at and after the critical threshold $t_c$ and denote them by $S_c^-=Ns(t_c)$ and $S_{c}^+=Ns(t_c+\Delta t)$, respectively. All these defined quantities i.e., $\Delta$, $t_c$, $S_c^-$ and $S_c^+$, are in fact random variables whose values are sample dependent. Therefore, our observables in this paper are the probability distribution function, the average and the root mean square of these variables. We consider $M=10^4$ number of independent realizations for every given system size $L$. We will then start our analysis with a set of measured variables as
\begin{equation}\label{2} 
Y = \{Y_1, Y_2, ..., Y_M\}_{Y =\, \Delta,\, t_c,\, S_c^-\mathrm{and}\, S_c^+},
\end{equation}
with the average
\begin{equation}\label{3} 
\bar{Y}(L)= \langle Y \rangle= \dfrac{1}{M}\sum_{i=1}^{M}\,\,Y_i,
\end{equation}
and the root mean square
\begin{equation}\label{4} 
\chi _{Y}(L) =\sqrt{\langle Y_i^2 \rangle}.
\end{equation}

\begin{figure}[t]
	\centering
	\includegraphics[width=3.4in]{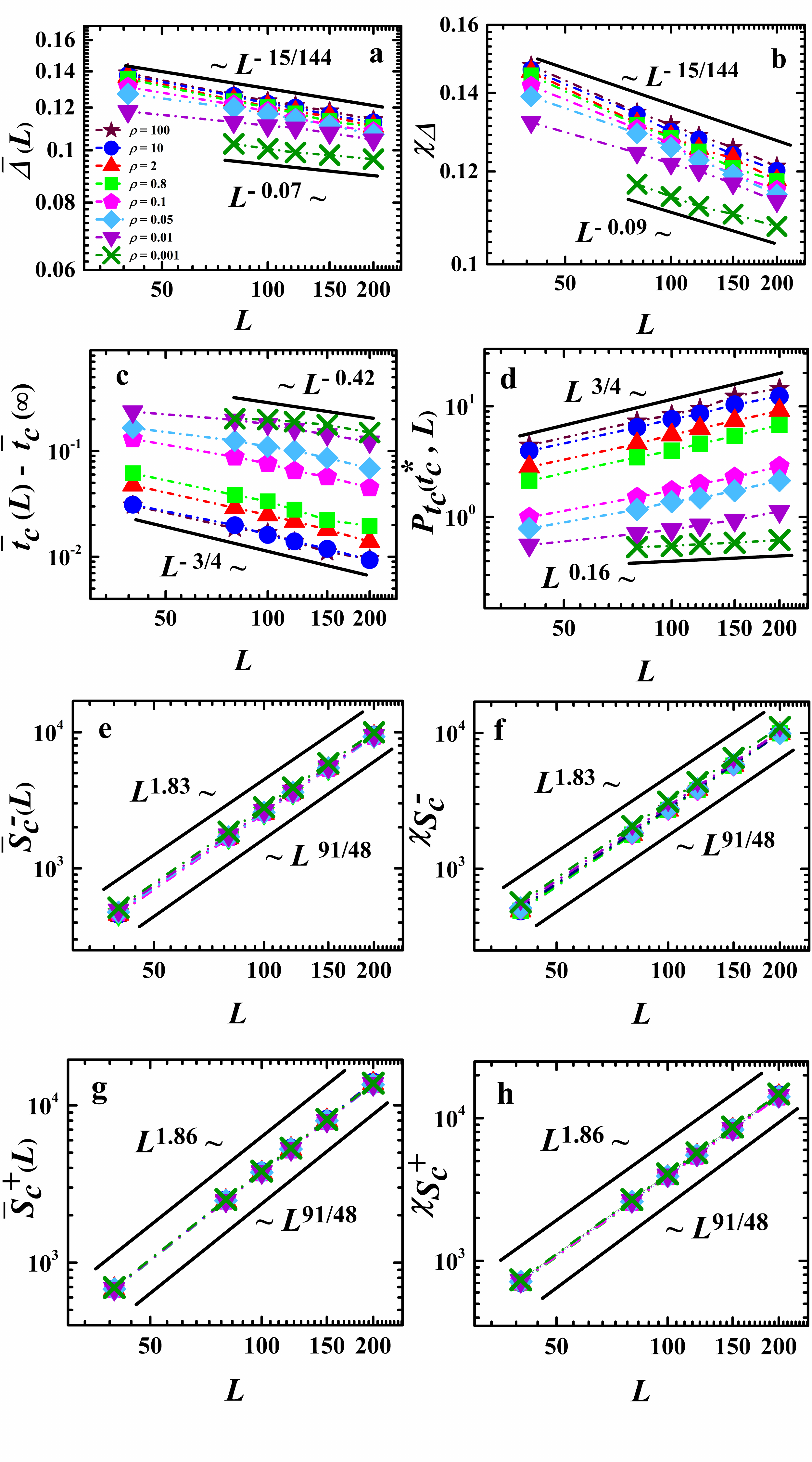}
	\caption{Plots for the computation of critical gap exponents and the fractal dimensions for a wide range of nonlocal link densities $\rho = 10^{-3}, 10^{-2},5\times 10^{-2}, 10^{-1}, 8\times 10^{-1}, 2, 10$, and $10^2$. The average size of the largest gap $\bar{\Delta}$ (a) and its fluctuation $\chi_\Delta$ (b), $\bar{t}_c(L)-t_c(\infty)$ (c) and $P_t(t_c^*, L)$ (d) as functions of $L$ give the exponents $\beta_1$, $\beta_2$, $\nu_1$ and $\nu_2$, respectively. The average size of the largest clusters $\bar{S}^-$ (e) and $\bar{S}^+$ (g) and their corresponding fluctuations $\chi_ {S^-}$ (f) and $\chi_ {S^+}$ (h) as functions of $L$ to extract the fractal dimensions. One of the solid lines show the best fit to our data for $\rho=10^{-3}$ (cross symbols) and the other shows comparison with the exact predictions available for the standard 2D percolation which is believed to explain the limit $\rho\rightarrow\infty$ of our model.  }
	\label{fig-2}
\end{figure}

\subsection{Critical gap exponents}

Finite-size scaling hypothesis provides a scaling framework for our observables as a function of the system size $L$, giving rise to a number of scaling and critical exponents which are able to classify the model \cite{r25, r26}. Various gap exponents and fractal dimensions can then be defined by the following proposed scaling behaviors,   

\begin{equation}\label{5} 
\bar{\Delta}\,(L) \sim L^{-\beta_1},\,\,\,\,\,\chi _{\Delta} \sim  L^{-\beta_2},
\end{equation}
\begin{equation}\label{6} 
\bar{S}{_c^-}(L) \sim L^{d_{f_1}^-},\,\,\,\,\,\chi_ {{S}{_c^-}} \sim  L^{d_{f_2}^-},
\end{equation}
\begin{equation}\label{7} 
\bar{S}{_c^+}(L) \sim L^{d_{f_1}^+},\,\,\,\,\,\chi _{{S}{_c^+}} \sim  L^{d_{f_2}^+},
\end{equation}
\begin{equation}\label{8} 
 \bar{t}{_c}(L)-\bar{t}{_c}(\infty) \sim L^{-1/\nu_1},\,\,\,\,\,\chi _{t_c} \sim  L^{-1/\nu_2}.
\end{equation}

The vanishing exponent $\beta_1=0$ can be used to determine if a percolation transition is discontinuous. However, models with $0<\beta_1<1$ can be either
continuous or discontinuous \cite{r26-1}. 
It can be also shown that $\beta_1=\beta_2$,  and the critical largest clusters share the same fractal dimensions: $d_{f_1}^-=d_{f_1}^+=d_{f_2}^-=d_{f_2}^+$. Moreover, the scaling relation $\beta_1=d-d_{f_1}$ holds for the gap exponents, which is similar to the known relation $\beta/\nu=d-d_{f}$ in terms of the standard order parameter and correlation exponents i.e., $\beta$ and $\nu$, respectively \cite{r26}. $\bar{t}_c(\infty)$ in Eq. (\ref{8}) denotes for the average critical threshold of the model in the thermodynamic limit where the system size goes to infinity i.e., in the limit $1/L\rightarrow 0$.

\begin{table*}[t]
	\centering
	\caption{The summarized asymptotic critical thresholds $\bar{t}_c(\infty)$, the effective critical gap exponents $\nu_1$, $\nu_2$, $\beta_1$, $\beta_2$, the fractal dimensions $d_{f_1}^\pm$, $d_{f_2}^\pm$, and the effective Fisher exponent $\tau$, estimated for various percentage $\rho$ of nonlocal links. The last row presents the same quantities known for the standard 2D site-percolation model.}
	\begin{tabular}{cccccccccccc}\hline\hline
		$\,\,\,\,\,\,\,\,\,\,\rho\,\,\,\,\,\,\,\,\,\,$&$\,\,\,\,\,\,\,\,\,\,\bar{t}_c(\infty)$\,\,\,\,\,\,\,\,\,\,\,\,\,\,\,&\,\,$1/\nu_1$\,\,\,\,\,\,\,\,\,\,\,\,\,\,\,&\,\,\,\, $1/\nu_2$\,\,\,\,\,\,\,\,\,\,\,\,\,\,\,&$\beta_1$\,\,\,\,\,\,\,\,\,\,\,\,\,\,\,&$\beta_2$\,\,\,\,\,\,\,\,\,\,\,\,\,\,\,&$d_{f_1}^-$\,\,\,\,\,\,\,\,\,\,\,\,\,\,\,& $d_{f_1}^+$\,\,\,\,\,\,\,\,\,\,\,\,\,\,\,&$d_{f_2}^-$\,\,\,\,\,\,\,\,\,\,\,\,\,\,\,& $d_{f_2}^+$\,\,\,\,\,\,\,\,\,\,\,\,\,\,\,&$\tau$& \\\hline
		0.001 & 1.50(5)& 0.42(3) &  0.16(3) & 0.07(2)  & 0.09(2) & 1.83(1) & 1.86(1) & 1.83(1) & 1.86(1) & 1.03(5)\\
		0.01 & 1.35(5) & 0.58(3) &  0.43(3) & 0.07(2)  & 0.09(2) &1.83(1) & 1.86(1) & 1.83(1) & 1.86(1) & 1.83(5)\\
		0.05 & 1.230(5) & 0.66(3) &  0.61(3) & 0.11(2) & 0.12(2) & 1.84(1) & 1.86(1) & 1.84(1) & 1.86(1) & 1.95(3)\\
		0.1 & 1.170(5)  & 0.66(3) & 0.65(3) & 0.12(2) & 0.13(2) & 1.85(1) & 1.86(1) & 1.85(1) & 1.86(1)& 1.99(3)\\
		0.8 & 1.025(5)  & 0.74(2) & 0.74(2) &  0.13(2)& 0.13(2) & 1.86(1) & 1.86(1) & 1.86(1) & 1.87(1) & 2.03(2)\\
		2 & 0.977(5)  & 0.75(1)  &  0.75(1) & 0.13(2) & 0.13(2) & 1.87(1) & 1.87(1) & 1.87(1) & 1.87(1) & 2.04(2) \\
		10  & 0.928(5)  & 0.75(1)  &  0.75(1)& 0.12(2) &0.12(2) & 1.88(1) &1.88(1) & 1.88(1) & 1.88(1) & 2.052(5)\\
		100  & 0.901(5)  & 0.75(1)  &  0.75(1) & 0.12(2) & 0.12(2) & 1.89(1) & 1.89(1) & 1.89(1) & 1.89(1) & 2.054(5)\\ \hline
		
		$2$D site-percolation&0.592746&3/4 & 3/4  & 5/48 & 5/48&91/48&91/48&91/48&91/48&187/91\\\hline
	\end{tabular}
	\label{table-1}
\end{table*}

\begin{figure}[t]
	\centering
	\includegraphics[width=3.0in]{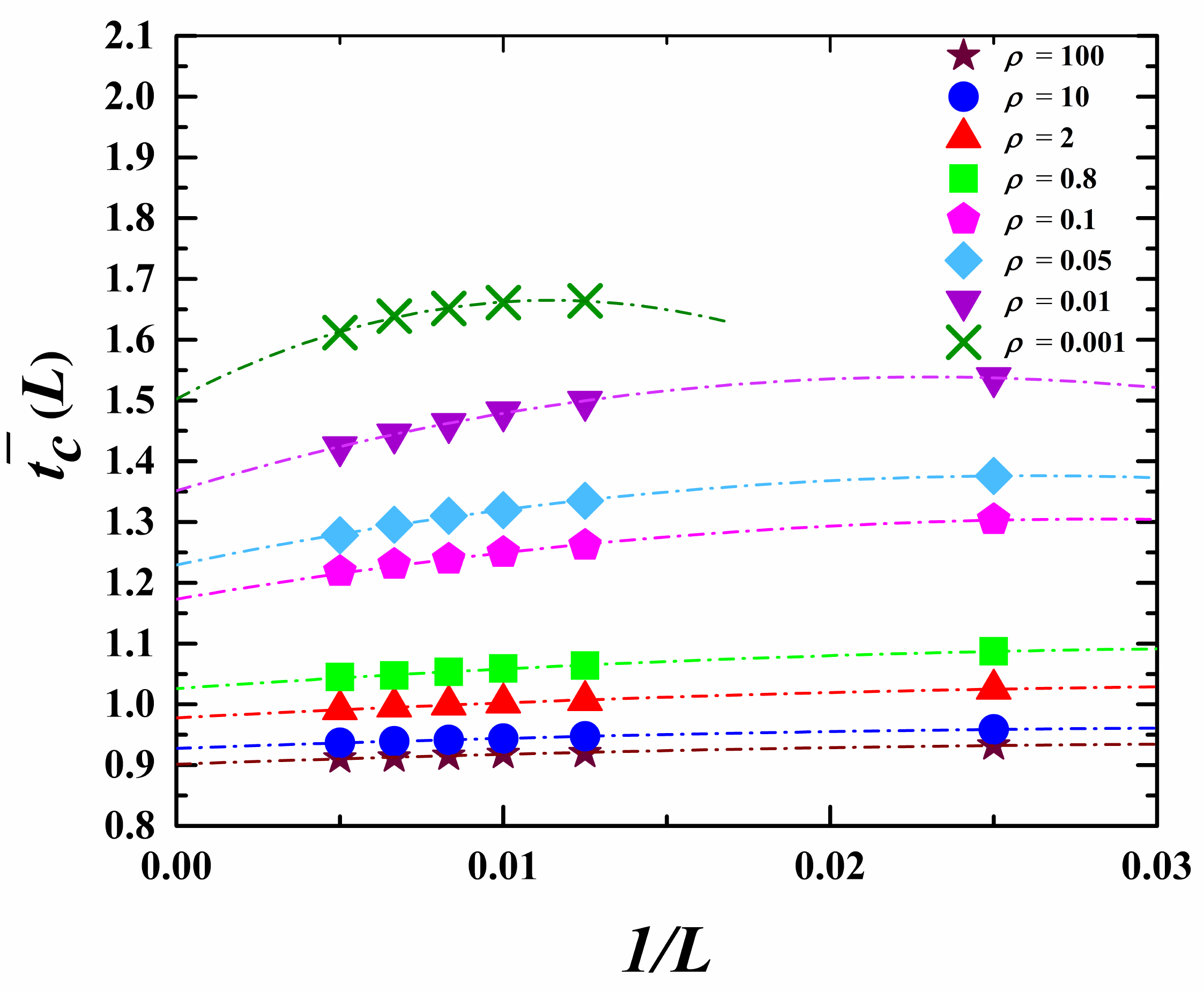}
	\caption{The average critical threshold $\bar{t}_c$ as a function of inverse size $1/L$ for $\rho = 10^{-3}, 10^{-2},5\times 10^{-2}, 10^{-1}, 8\times 10^{-1}, 2, 10$, and $10^2$. The dotted-dashed lines show our extrapolations to the infinite system size at $1/L\rightarrow 0$ to estimate the asymptotic $\bar{t}_c(\infty)$, listed in the first column of Table \ref{table-1}.}
	\label{fig-3}
\end{figure}

\subsection{Universal scaling functions}

One may hypothesize that the distributions $P_Y(Y, L)$ of the introduced random observables $Y=\Delta, S_c^{\pm}$ and $t_c$, for a system of size $L$, have to take the following finite-size scaling forms \cite{r26-2}
\begin{equation}\label{16} 
P_{\Delta}(\Delta, L) = L^{\beta_2}\,f_{\Delta}(\Delta\,L^{\beta_2}),
\end{equation}

\begin{equation}\label{17} 
P_{S^{\pm}_c}(S_c^{\pm}, L) = L^{-d_{f^{\pm}_2}}\,f_{S_c^{\pm}}(S_c^{\pm}\,L^{-d^{\pm}_{f_2}}),
\end{equation}

\begin{equation}\label{18} 
P_{t_c}(t_c, L) = L^{1/\nu_2}\,f_{t_c}(\delta t_c\,L^{1/\nu_2}),
\end{equation}
where $f_Y(\cdot)$ are the universal scaling functions, and $\delta t_c=t_c - \bar{t}_c(\infty)$. The size of the largest gap $\Delta$ in the order parameter, as the maximum of random variables drawn from independent realizations of the entire percolation process, and the corresponding size of the largest cluster $S^{\pm}_c$ which dominates the system at $t_c$, can be viewed as the extreme events whose distributions are predicted by the extreme-value theory \cite{r26-3}.
The critical threshold $t_c$, on the other hand, is a non-extremal variable drawn and averaged from independent realizations. Therefore, according to the central limit theorem, we expect a Gaussian distribution for $t_c$. 

The scaling forms (\ref{16})-(\ref{18}) also suggest an alternative approach to estimate the involved critical exponents by data collapse onto the universal scaling curves. In particular, in order to estimate the exponent $\nu_2$, we will use the following finite-size scaling law 
\begin{equation}\label{15} 
P_{t_c}(t_c^*, L) \sim L^{1/\nu_2},
\end{equation} 
where $t_c^*$ is the mode of the distribution $P_{t_c}(t_c, L)$ which is also very close to $\bar{t}_c(\infty)$ due to the Gaussian form of the distribution.

\begin{figure}[b]
	\centering
	\includegraphics[width=3.4in]{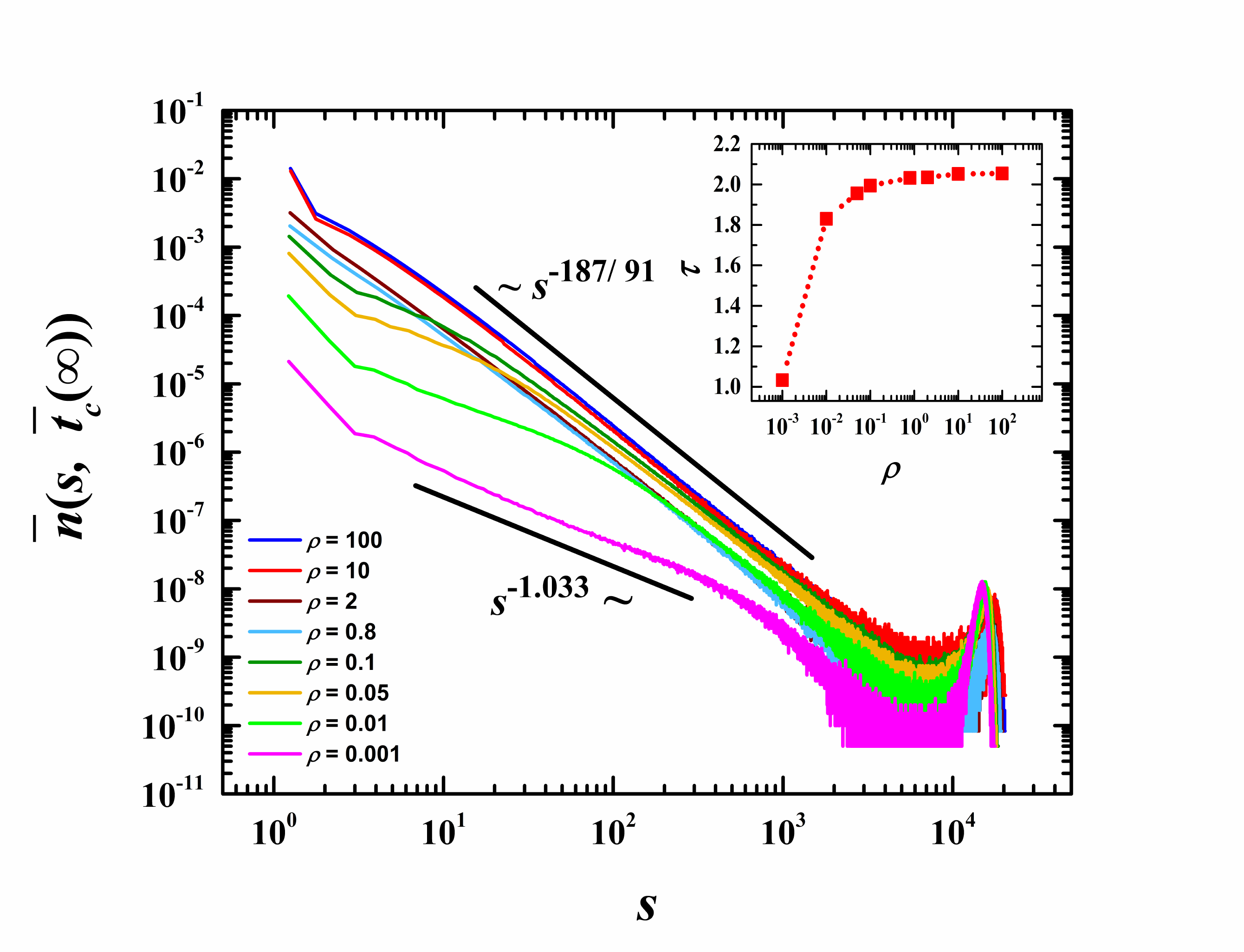}
	\caption{Estimation of the Fisher exponent $\tau$ at the asymptotic critical threshold for $\rho = 10^{-3}, 10^{-2},5\times 10^{-2}, 10^{-1}, 8\times 10^{-1}, 2, 10$, and $10^2$, extracted from the scaling region of the average number $\bar{n}(s, t_c(\infty))$ of clusters of size $s$. The averages are taken over $10^6$ independent realizations for a system of linear size $L=200$. The lower solid line shows the best fit to our data for $\rho=10^{-3}$, and the upper solid line compares our data for the largest $\rho$, with the predicted behavior for the standard 2D percolation model. }
	\label{fig-4}
\end{figure}

\begin{figure}[h!]
	\centering
	\includegraphics[width=3.2in]{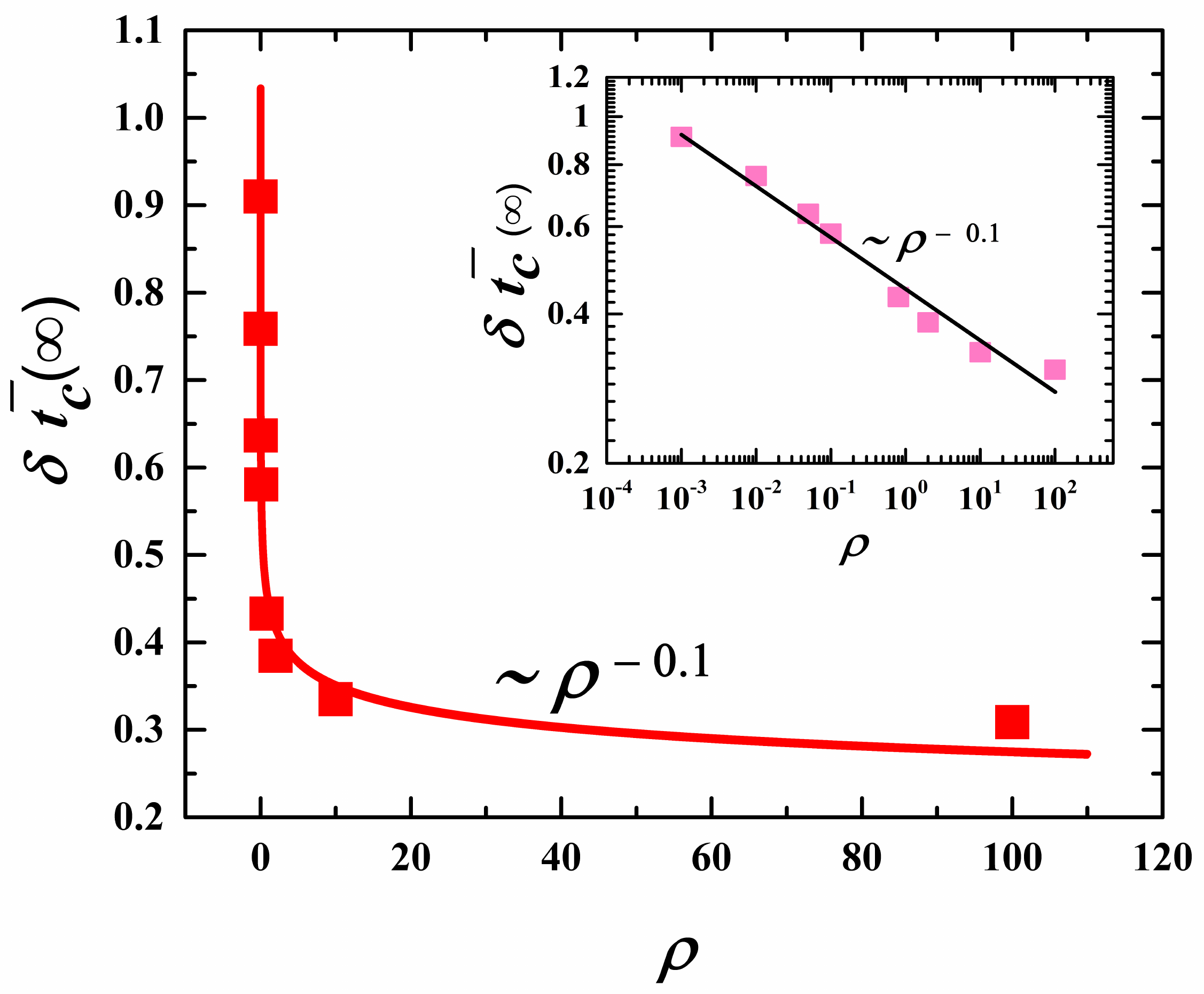}
	\caption{The difference time threshold $\delta\bar{t}_c(\infty)=\bar{t}_c(\infty)-\bar{t}^{\text{site}}_c(\infty)$ as a function of $\rho$. The solid line shows the best power law fit $\propto \rho^{-\zeta}$ to our data with the exponent $\zeta=0.10(2)$ estimated from the log-log plot of the data shown in the Inset. }
	\label{fig-5}
\end{figure}

\begin{figure}[t]
	\centering
	\includegraphics[width=3.4in]{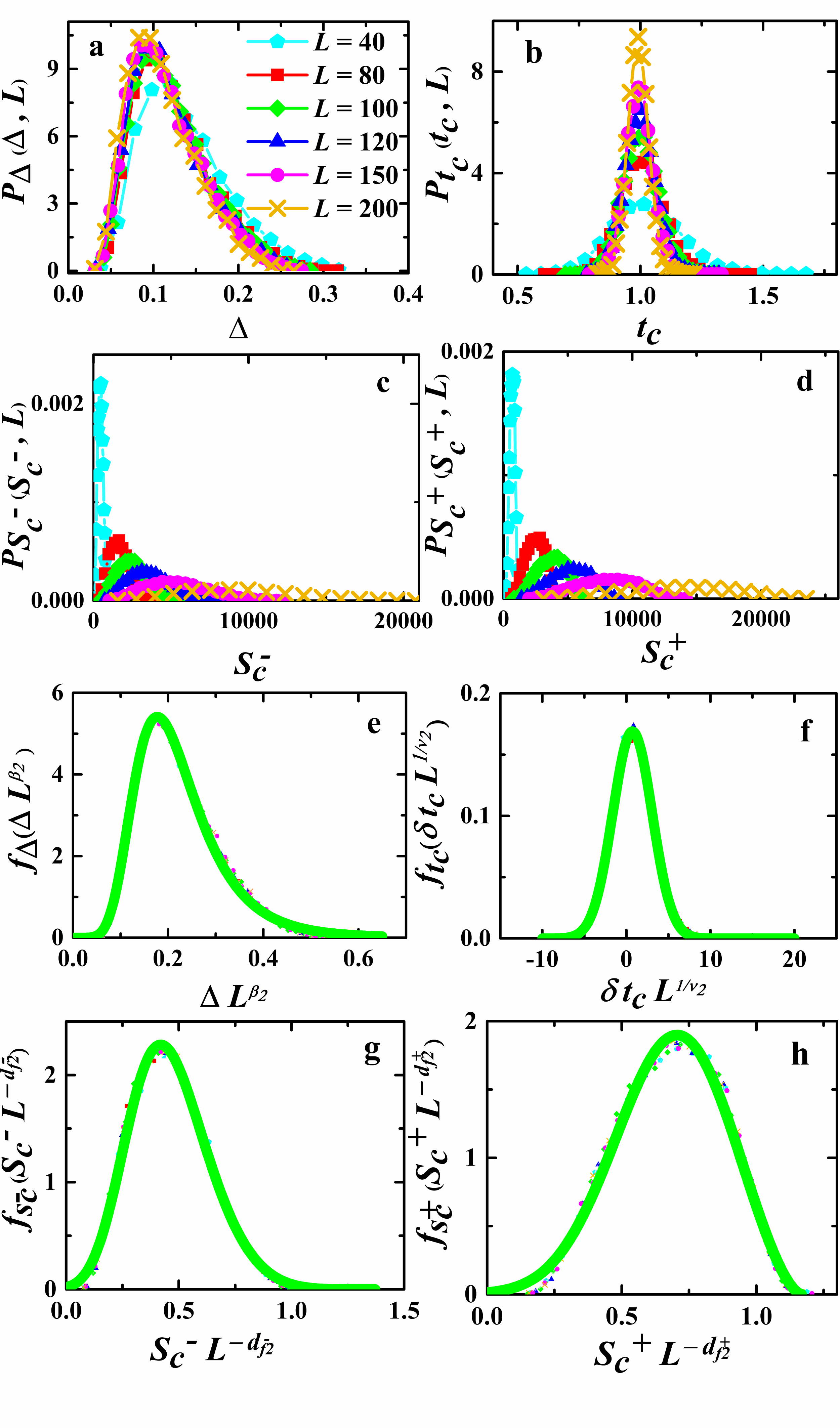}
	\caption{Universal scaling functions for $\rho = 2$. (a–d) Distributions of the observables $\Delta$, $t_c$, $S_c^-$ and $S_c^+$ for different system sizes $L=40, 80, 100, 120, 150$ and $200$. (e–h) Corresponding universal scaling functions of rescaled variables $\Delta L^{\beta_2}$, $\delta t_c L^{1/\nu_2}$, $S_c^- L^{-d^-_{f_2}}$ and $S_c^+L^{-d^+_{f_2}}$. The green solid lines are the best fit to our data compatible with Frechet extreme value in Eq. \ref{GEV} with $\xi=+0.064(5)$, $\mu=0.181(5)$ and $\sigma=0.069(5)$ (e), Gaussian distribution (f), and Weibull extreme value with $\xi=-166(5)$, $\mu=0.0.390(5)$ and $\sigma=0.163(5)$ in (g), and $\xi=-0.360(5)$, $\mu=0.616(5)$ and $\sigma=0.212(5)$ in (h).}
	\label{fig-6}
\end{figure}

\section{Numerical results} \label{numerics}

In this Section we are going to present the results of our numerical simulations of our model previously described in Section \ref{Model}. We have carried out extensive simulations for different linear sizes $L$ and various choices of $\rho$ over a range of $5$ orders of magnitude $\rho\in[10^{-3}, 10^2]$ to best estimate the critical exponents and threshold for every $\rho$  in the asymptotic limit $L\rightarrow\infty$. Figures \ref{fig-2}a,b demonstrate the plots  of  $\bar{\Delta}(L)$ and  $\chi_{\Delta}$ as a function of $L$ for various choices of $\rho$. We use logarithmic scales on both the horizontal and vertical axes to extract the critical exponents in all graphs. The exponents $\beta_1$ and $\beta_2$ are measured by examining the scaling relations defined in Eq. (\ref{5}). As shown in the figures (and summarized in Table \ref{table-1}) we find a spectrum of exponents which vary with $\rho$. The lower solid lines show the best fit to our data for $\rho=10^{-3}$ giving $\beta_1\sim 0.07(1)$ and $\beta_2\sim 0.09(1)$. The upper solid lines in the graphs show the corresponding exact exponents predicted for 2D standard percolation model.\\
In the limit $\rho\rightarrow\infty$, there are infinitely many nonlocal links which makes the RW's movement totally nonlocal and the model effectively becomes equivalent to a 2D site-percolation model in which every site is randomly chosen and gets infected. Therefore, we expect that in the large $\rho$ limit, our model falls into the 2D percolation universality class with exactly known critical exponents. As can be seen in the figures (and Table \ref{table-1}), the agreement with our expectation is evident for $\rho\ge 1$. \\In order to measure the exponents $\nu_1$ and $\nu_2$ by using Eqs. (\ref{8}) and (\ref{15}), we first estimate the asymptotic average critical threshold $\bar{t}_c(\infty)$ for every $\rho$ by plotting our data for the average value of the threshold $\bar{t}_c(L)$ obtained for a given system size $L$ as a function of $1/L$ and extrapolating to the infinite system size limit $1/L\rightarrow 0$ (see Fig. \ref{fig-3} and second column in Table \ref{table-1}). Figures \ref{fig-2}c,d show the plots $\bar{t}_c(L)-\bar{t}_c(\infty)$ and $P_t(t^*_c, L)$ as a function of $L$, yielding the exponents $\nu_1$ and $\nu_2$, respectively, also summarized in Table \ref{table-1}. We find $1/\nu_1=1/\nu_2\sim 0.75(1)$ for $\rho>1$, which agrees very well with the exact exponent $1/\nu=3/4$ for 2D percolation. However, for smaller values of $\rho$ we find that $\nu_1\ne\nu_2$ whose values increase as $\rho$ decreases (see the third and forth columns in Table \ref{table-1}). This finding is intuitively understandable, since it is consistent with the known observation that the correlation exponent increases with increase in the underlying dimensionality, where, in our case, the decrease in $\rho$ decreases the effective dimensionality from infinity (for $\rho\rightarrow\infty$) to a pure 2D (for $\rho=0$).\\We have also measured the fractal dimensions defined in Eqs. (\ref{6}) and (\ref{7}) for the giant cluster of infected individuals at the critical threshold which are shown in Figs. \ref{fig-2}e-h. For higher values of $\rho$, the sites all over the lattice become more accessible for the random walker to be visited and, thus, leads to  formation of more compact clusters with higher fractal dimension (Note that despite the local dependencies introduced by consecutive steps of an ordinary RW, the overall distribution of the sites visited by the RW in the presence of the nonlocal extra links is well spread across the system.)

The other geometric exponent that characterizes the universality class of critical systems is provided by the distribution of cluster sizes which appears to follow a power law \begin{equation}
\bar{n}(s, \bar{t}_c(\infty))\sim s^{-\tau},
\end{equation} for large $s$, where $\bar{n}$ gives the average number of clusters of $s$ connected sites, per lattice site, measured at the asymptotic critical point $\bar{t}_c(\infty)$ for each $\rho$. The Fisher exponent $\tau$ is a universal quantity whose value is the same for all systems of a given class. \\ Figure \ref{fig-4} shows our computations for $\bar{n}(s, \bar{t}_c(\infty))$ for various values of $\rho$ on a system of linear size $L=200$. The estimated values of Fisher exponents in the scaling region are presented in the Inset of Fig. \ref{fig-4} as a function of $\rho$. The distribution is proportional to the inverse cluster size $\bar{n}(s, \bar{t}_c(\infty))\propto 1/s$ with the exponent $\tau\sim 1$ for the smallest $\rho (=10^{-3})$, while the exponent converges to the exact value $\tau=187/91$, known for the standard 2D percolation, in the limit $\rho\rightarrow\infty$. Our estimated Fisher exponents are summarized in the last column of Table \ref{table-1}.

A closer look at Figure \ref{fig-4} shows that for each intermediate $\rho$, the corresponding graph shows a crossover behavior between two regimes with small ($s<s_c$) and large ($s>s_c$) cluster sizes, where the crossover point $s_c$ vanishes as $\rho$ increases. The behavior for $s<s_c$ is governed by the RWs on a pure square lattice with $\tau\sim1$ and, for $s>s_c$, it is given by $\tau=187/91$ for the ordinary percolation in two dimensions. This observation strongly suggests that the dependence on $\rho$ in our estimated exponents listed in Table \ref{table-1}, is just a reminiscence of a crossover behavior that system undergoes between the 2D percolation and 2D RWs universality classes. This is however, due to the finite-size of the system, which due to its high computational cost, we are not able to perform simulations for much larger systems. This is why we call our estimated exponents "effective exponents".

A remarkable aspect of our model is that it allows us to define and predict the emergence of a pandemic in a more precise form as the state of an epidemics where the percolation threshold $\bar{t}_c(\infty)$ is reached. This gives the time at which an epidemic outbreak becomes a pandemic whose prediction is of great importance in epidemiology. In our model, the temporal and spatial progress of a global epidemic is totally governed by the percentage of available nonlocal links in the sample space controlled by the parameter $\rho$. In the limit $\rho\rightarrow\infty$, the threshold is given by the critical point of a 2D site-percolation at $\bar{t}^{\text{site}}_c(\infty) \sim 0.5927$. We find that the difference time threshold $\delta\bar{t}_c(\infty)=\bar{t}_c(\infty)-\bar{t}^{\text{site}}_c(\infty)$ exhibits a scaling relation with $\rho$ as follows
\begin{equation}
\delta\bar{t}_c(\infty)\sim \rho^{-\zeta},
\end{equation} with the exponent $\zeta\sim 0.10(2)$ (see Fig. \ref{fig-5}). This relationship suggests that the onset of a global pandemic can be algebraically delayed by reducing the percentage of nonlocal links in our setting.

\begin{figure}[t]
	\centering
	\includegraphics[width=3.1in]{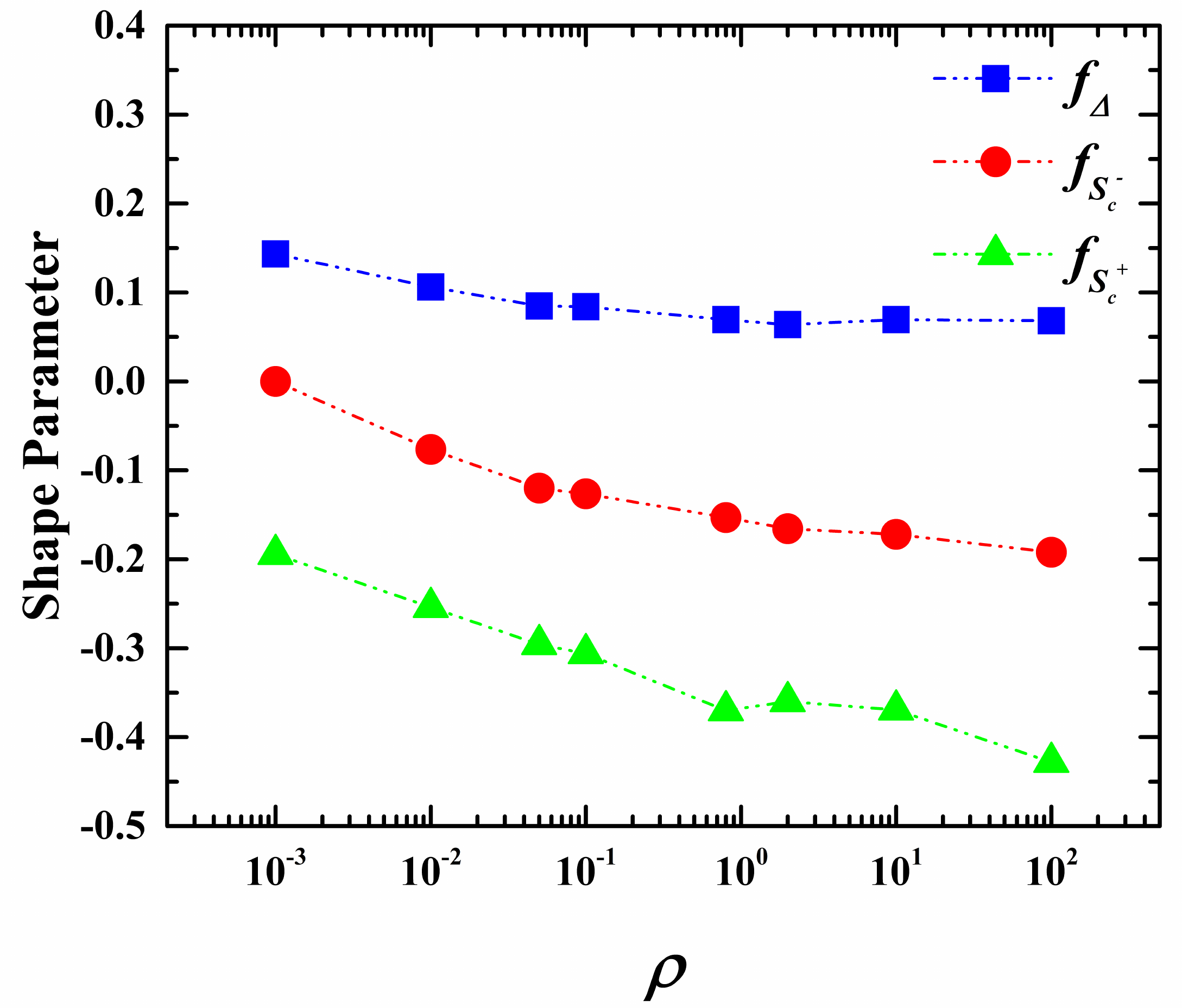}
	\caption{Shape parameters $\xi$ defined in the Generalized Extreme Value (GEV) distribution (Eq. \ref{GEV}), as functions of $\rho$ obtained from the best fits to our data for the universal scaling functions of the observables $\Delta$ (blue squares), $S_c^-$ (red circles) and $S_c^+$ (green triangles). $f_\Delta$ is compatible with Frechet distribution for all $\rho$, $f_{S_c^-}$ with Weibull distribution for all $\rho$ expect for $\rho=10^{-3}$ which is more close to the Gumbel distribution with $\xi=0$. $f_{S_c^+}$ is in agreement with the Weibull distribution for all $\rho$ with negative shape parameters $\xi<0$. }
	\label{fig-7}
\end{figure}

In the rest of the present Section, we are going to establish a connection between the universal scaling functions $f_Y(\cdot)$ mentioned in Eqs. (\ref{16})-(\ref{18}) with the extreme value theory. This connection was first introduced in \cite{r26} for a wide class of percolation problems in different dimensions where $f_Y(\cdot)$ for $Y= \Delta-\bar{\Delta}(L)$ and $S_c^--\bar{S}_c^-(L)$ were shown to be consistent with the Gumbel distribution \cite{r29}. Here by $Y$ we do not mean the fluctuation of the corresponding random variables about their average, but we simply mean the variables themselves i.e., $Y$ denotes for either $\Delta$, $S_c^-$ or $S^+_c$ directly. This introduces computational simplicity, in particular for problems in which the average of the variables is not in a priori known.

In order to obtain the universal scaling functions $f_Y(\cdot)$ defined in Eqs. (\ref{16})-(\ref{18}), we first compute the distributions of corresponding variables i.e., $Y=\Delta, S_c^{\pm}$ and $t_c$, for various system sizes. Figures \ref{fig-6}a-d. show an example of the distributions for $\rho = 2$ and system sizes $L = 40, 80, 100, 120, 150$ and $200$. Strong size-dependence is evident in these figures. As shown in subsequent figures \ref{fig-6}e-h, when the horizontal and vertical axes in Figs. \ref{fig-6}a-d are suitably rescaled, all data collapse onto a single universal curve. We find excellent data collapse over the entire range of scaling variables. The best fits to our data are shown with the solid (green) lines in Figs. \ref{fig-6}e-h. We find that the universal function for $Y=t_c$, is  in perfect agreement with a Gaussian function with R-square $R^2_{f_{t_c}}=0.9977(10)$ (see Fig. \ref{fig-6}f). For other variables $Y=\Delta, S_c^{\pm}$, we have examined the following Generalized Extreme Value (GEV) distribution which unifies the types I, II, and III extreme value distributions into a single family, by allowing a continuous range of possible shapes\begin{align}\label{GEV} 
f(z) =  \frac{1}{\sigma}\,\,t(z)^{\xi+1}\,\,e^{-t(z)}
\end{align} with
\begin{align}\nonumber
t(z) =\left\{ \begin{array}{rcl}
\Big(1+\xi\,\,(\frac{z-\mu}{\sigma})\Big)^{-1/\xi}\hspace{0.8cm}  &  \text{if} & \xi \neq 0,\\
\exp\big(-\frac{(z-\mu)}{\sigma}\big) \hspace{1.35cm} & \text{if} & \xi = 0.\\
\end{array}\right.
\end{align}
It is parameterized with location $\mu$, scale parameter $\sigma$, and a shape parameter $\xi$. When $\xi< 0$, the GEV is equivalent to the Weibull (type III) extreme value. When $\xi> 0$, the GEV is equivalent to the Frechet (type II), and in the limit $\xi\rightarrow 0$, it becomes the Gumbel (type I) distribution. The best fits in MATLAB to the universal scaling functions
$f_{\Delta}(\Delta\,L^{\beta_2})$ (Fig. \ref{fig-6}e),  $f_{S^-_c}(S_c^-\,L^{-d_{f2}^-})$ (Fig. \ref{fig-6}g) and $f_{S^+_c}(S_c^+\,L^{-d_{f2}^+})$ (Fig. \ref{fig-6}h), are attained with the shape parameters $\xi=+0.064(5)$ ($R^2_{f_{\Delta}}=0.9958(10)$)
, $-0.166(5)$ ($R^2_{f_{S_c^-}}=0.9927(10)$) and $ -0.360(5)$ ($R^2_{f_{S^+}}=0.9827(10)$), respectively. Our results indicate that the universal function for the maximum jumps $\Delta$ is compatible with the Frechet distribution and the ones for $S_c^\pm$ can be well described by Weibull distributions. The shape parameters as a function of $\rho$ for $f_{\Delta}$, $f_{S_c^-}$ and $f_{S_c^+}$ are presented in figure \ref{fig-7}. Despite a slight change, almost the same behavior can be observed for the whole range of studied $\rho$.

\section{Concluding remarks} \label{concl}
We have introduced a variant of percolation model to study the effect of the underlying network topology, controlled by the density $\rho$ of nonlocal links, on the spread of an infection/information when an infected/informed person performs independent random walks on a graph embedded in two dimensions. Our model reduces to the ordinary nonequilibrium diffusion problem in 2D in the limit $\rho\rightarrow 0$ and, to the standard site-percolation model in pure 2D in the limit $\rho\rightarrow\infty$ with a known equilibrium geometric phase transition. Due to finite-size effects, we find a spectrum of effective universality classes in terms of the single control parameter $\rho$, crossing over between these two important prototype models in the statistical physics. We have carried out extensive numerical simulations to estimate the effective exponents from a finite-size scaling analysis of the extreme event records in the temporal evolution of the largest cluster in an ensemble of independent realizations. This approach facilitates the practical applications of our model into the realistic situations. We find that the time threshold at which the onset of a pandemic occurs, exhibits a scaling relation with the control parameter $\rho$ that is a quantity of great interest in the epidemiology. We have also established a connection between the extreme value theory and the universal scaling functions describing the probability distribution of the extreme observables. Our study calls for further future research to mathematical understanding of our model to realize the nature of the transitions as well as exact description of the crossover behavior in terms of the density of the nonlocal links.   
\\[\baselineskip]

\section*{Acknowledgments} A.A.S. acknowledges the supports from the Alexander von Humboldt Foundation (DE) and the research council of the University of Tehran. We also thank the High Performance Computing (HPC) center in the University of Cologne, Germany,
where a part of computations have been carried out. M.F. would like to thank M. Ebrahimi for her helps on programming.

\end{document}